\documentclass[conference]{IEEEtran}
\IEEEoverridecommandlockouts
\usepackage{cite}
\usepackage{amsmath,amssymb,amsfonts}
\usepackage{algorithm}
\usepackage{algorithmic}
\usepackage{graphicx}
\usepackage{textcomp}
\usepackage{xcolor}

\usepackage{algorithm}
\usepackage{wrapfig}
\usepackage{booktabs}
\usepackage{multirow}
\usepackage{color}
\usepackage{hyperref}
\def\BibTeX{{\rm B\kern-.05em{\sc i\kern-.025em b}\kern-.08em
    T\kern-.1667em\lower.7ex\hbox{E}\kern-.125emX}}
    
\newcommand{\modified}[1]{{\color{black} #1}}

\begin{document}

\title{GE-SpMM: General-purpose Sparse Matrix-Matrix Multiplication on GPUs for Graph Neural Networks
}

\author{Guyue~Huang,
        Guohao~Dai,
        Yu~Wang
        and~Huazhong~Yang
        \thanks{All authors are with Department of Electronic Engineering, Tsinghua University, Beijing, China (email: huang-gy16@mails.tsinghua.edu.cn, daiguohao@mail.tsinghua.edu.cn, \{yu-wang, yanghz\}@tsinghua.edu.cn).}
}

\maketitle

\begin{abstract}
Graph Neural Networks (GNNs) based graph learning algorithms have achieved significant improvements in various domains. Sparse Matrix-Matrix multiplication (SpMM) is a fundamental operator in GNNs, which performs a multiplication operation between a sparse matrix and a dense matrix. 
Accelerating SpMM on parallel hardware like GPUs can face the following challenges:
From the GNN application perspective, the compatibility needs to be considered. General GNN algorithms require SpMM-like operations (e.g., pooling) between matrices, which are not supported in current high-performance GPU libraries (e.g., Nvidia cuSPARSE~\cite{cusparse}). Moreover, the sophisticated preprocessing in previous implementations will lead to heavy data format conversion overheads in GNN frameworks. From the GPU hardware perspective, optimizations in SpMV (Sparse Matrix-Vector) designs on GPUs do not apply well to SpMM. SpMM exposes the column-wise parallelism in the dense output matrix, but straightforward generalization from SpMV leads to inefficient, uncoalesced access to the GPU global memory. Moreover, the sparse row data can be reused among GPU threads, which is neither possible in SpMM designs inherited from SpMV. 

To tackle these challenges, we propose GE-SpMM\footnote{The project is open-sourced at \url{https://github.com/hgyhungry/ge-spmm}}. GE-SpMM performs SpMM-like operation on sparse matrices represented in the most common Compressed Sparse Row (CSR) format. Thus, GE-SpMM can be efficiently embedded in GNN frameworks with no preprocessing overheads and support general GNN algorithms. We introduce the Coalesced Row Caching method to process columns in parallel and ensure efficient coalesced access to the GPU global memory. We also present the Coarse-grained Warp Merging method to reduce redundant data loading among GPU warps. Experiments on a real-world graph dataset show that GE-SpMM achieves up to 1.41$\times$ speedup over Nvidia cuSPARSE~\cite{cusparse} and up to 1.81$\times$ over GraphBLAST~\cite{design}. 
We also embed GE-SpMM in GNN frameworks and get up to 3.67$\times$ speedup over popular GNN models like GCN~\cite{gcn} and GraphSAGE~\cite{sage}.

\end{abstract}


\section{Introduction} \label{sec:intro}
Machine learning algorithms on graphs, especially the recently proposed Graph Neural Networks (GNNs), have been successfully applied to tasks such as link prediction and node classification \cite{gcn,sage,rgcn,gin,fout2017protein,wu2020comprehensive}. Acceleration of GNN systems is crucial to solving real-world problems because of days of execution time (e.g., it takes up to 78 hours to train a GNN model with 7.5 billion edges on 16 GPUs~\cite{ying2018graph}). In GNN algorithms, data are organized in the graph structure composed of vertices (nodes) and edges (links). Each vertex in the graph is associated with a feature vector, and these feature vectors are propagated through edges to perform GNN algorithms. Thus, aggregating the feature vectors of neighbor vertices is a fundamental operation in GNNs, which can be expressed with Equation~(\ref{equ:reduce}). 

\begin{equation}\label{equ:reduce}
    \centering
    \overrightarrow{f_u} = reduce\_op(\{\overrightarrow{f_v}\}), \exists edge_{v\rightarrow u}
\end{equation}

\begin{figure}[!t]
\centering
\includegraphics[width=0.45\textwidth]{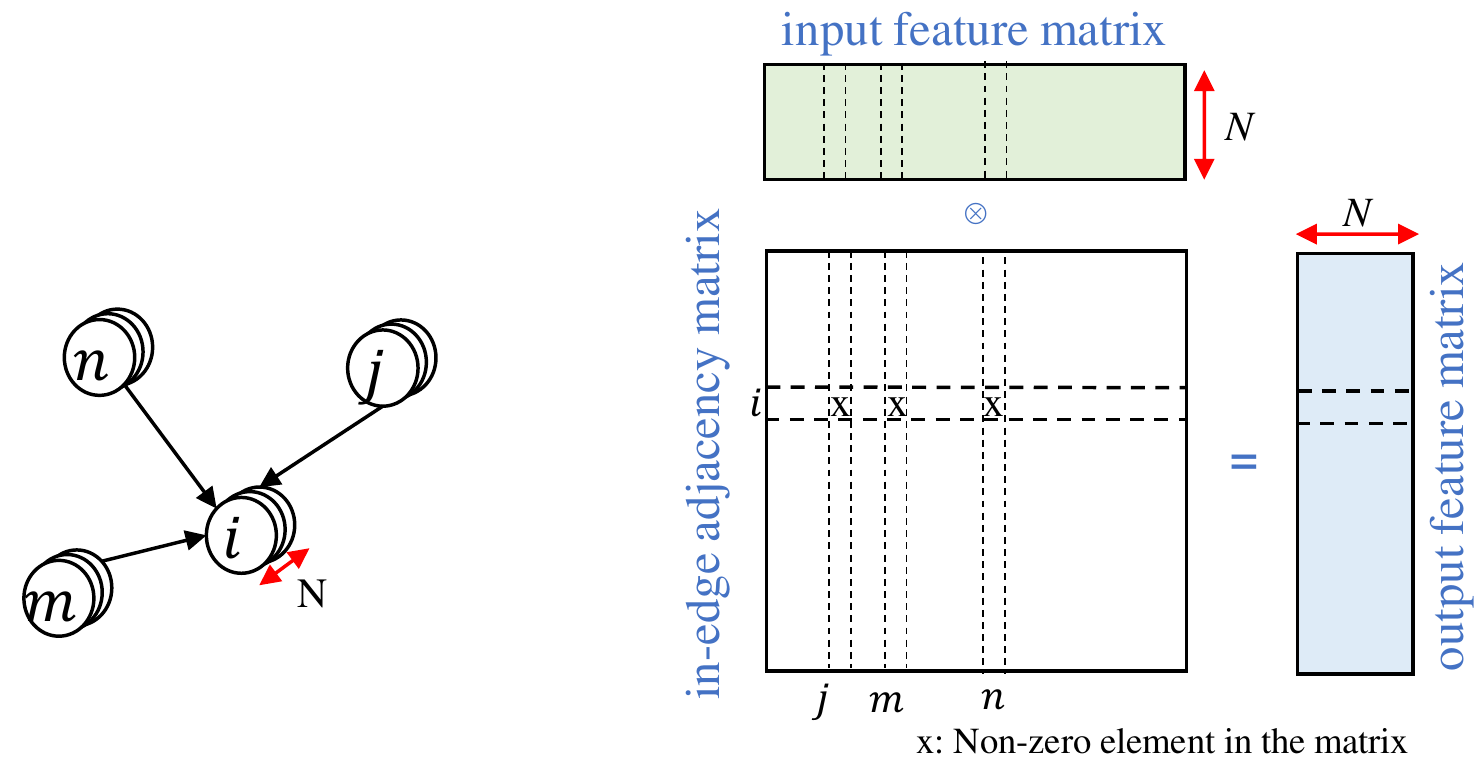}
\vspace{-5pt}
\caption{SpMM operation from a graph- and matrix-perspective. Left: aggregating feature vectors (length=$N$) from Vertex $j,m,n$ to Vertex $i$. Right: the SpMM operation between the in-edge adjacency matrix and the feature matrix, involving the $j$-th,$m$-th,$n$-th vectors in the feature matrix. The $\otimes$ operation represents the general-purpose aggregating operation.}
\label{fig:spmm-in-gnn}
\vspace{-15pt}
\end{figure}

In Equation~(\ref{equ:reduce}), $\overrightarrow{f_u}$ represents the feature vector of a vertex $u$ in the graph, while $\{\overrightarrow{f_v}\}$ denotes the collection of feature vectors of $u$'s in-edge neighbors. The $reduce\_op$ is the aggregation operation to generate a new feature vector of $u$, which varies in different GNN algorithms~\cite{gcn,sage}. The edges between vertices in natural graphs are usually sparse, and the graph structure can be represented using a sparse adjacency matrix. Fig.~\ref{fig:spmm-in-gnn} shows that the aggregation operation on the graph can be treated as a customized multiplication between the sparse adjacency matrix (representing the graph) and dense feature matrix (representing feature vectors of vertices). When the $reduce\_op$ in Equation~(\ref{equ:reduce}) is taking sum, the operation between the adjacency matrix and the feature matrix is a standard Sparse Matrix-Matrix Multiplication (SpMM). When a customized $reduce\_op$ is adopted (e.g., pooling), we call it a general SpMM-like operation.

Being a primary operation in GNN models, SpMM is also a time-consuming step even on parallel hardware like GPUs. We profile the percentage of SpMM operations\footnote{Percentage of CUDA time, reported by PyTorch~\cite{pytorch} autograd profiler.} during a GCN~\cite{gcn} training procedure. In GCN training, the forward and backward of graph convolution layers both involve SpMM. As listed in Table~\ref{tb:spmm-percentage}, SpMM operations take $\sim30\%$ of the total time in the example code provided by DGL~\cite{dgl} with default settings. \modified{Dense matrix multiplications take $\sim10\%$, and the rest of the operators all take less than 10\%.} Thus, accelerating SpMM operations in GNN frameworks is significant for improving performance. However, current SpMM acceleration solutions on GPUs still face challenges from 1) meeting GNN application requirements, and 2) full utilization of global memory bandwidth of the GPU hardware.

\begin{table}[h]
\vspace{-10pt}
\caption{Percentage of SpMM in CUDA~\cite{cuda} time during GCN training on GTX1080Ti}
\vspace{-5pt}
\label{tb:spmm-percentage}
    \centering
    \begin{tabular}{ c | c }
        \hline
         Graph & SpMM percentage \\
         \hline
         \verb|Cora|       & 33.1\% \\ 
         \verb|Citeseer|   & 29.3\% \\
         \verb|Pubmed|     & 29.8\% \\
         \hline
    \end{tabular}
\vspace{-5pt}
\end{table}

From the GNN application perspective, embedding SpMM designs in GNN frameworks has at least two requirements: support for general SpMM-like operators (rather than the standard SpMM), and no (or very low) data format conversion overhead in the whole framework. In terms of general SpMM-like operators, existing GNN frameworks cannot achieve as good performance as using standard SpMM in cuSPARSE~\cite{cusparse} library. Current SpMM researches claiming better performance than cuSPARSE rely on preprocessing sparse matrix and cannot be conveniently adopted. Existing GNN systems like DGL~\cite{dgl} relies on Nvidia cuSPARSE to perform standard SpMM, but falls back to its own implementation for SpMM-like operations because they are not provided in cuSPARSE. Table~\ref{tb:spmm-and-spmmlike} shows the comparison of the same aggregation step in two models:  GraphSAGE-GCN~\cite{sage}, which internally calls SpMM, and GraphSAGE-pool~\cite{sage}, which internally calls SpMM-like. We again use example codes provided by DGL with default parameters. The results show that current implementation of SpMM-like in DGL cannot compete with the performance of cuSPARSE. On the other hand, although recent studies on SpMM~\cite{rsspmm, aspt} in high-performance computing fields achieve even better performance than cuSPARSE, they cannot be directly adopted by GNN frameworks. These implementations require preprocessing on input sparse matrix, which is hard to be integrated into GNN frameworks. Also, the extra time spent on preprocessing cannot be compensated by SpMM performance gain if SpMM is performed only a few times in GNNs, as is the case in GNN direct inference or batched training. 

\begin{table}[h]
\vspace{-10pt}
\caption{SpMM and SpMM-like comparison in DGL~\cite{dgl} on GTX 1080Ti}
\vspace{-5pt}
\label{tb:spmm-and-spmmlike}
    \centering
    \begin{tabular}{c | c}
        \hline
         Graph & SpMM-like perf. loss against SpMM in GraphSAGE~\cite{sage}\\
         \hline
         \verb|Cora|&8.8\%\\ 
         \verb|Citeseer|&89.2\%\\
         \verb|Pubmed|&139.1\%\\
         \hline
    \end{tabular}
\vspace{-5pt}
\end{table}

\begin{figure}[t]
\centering
\includegraphics[width=0.45\textwidth]{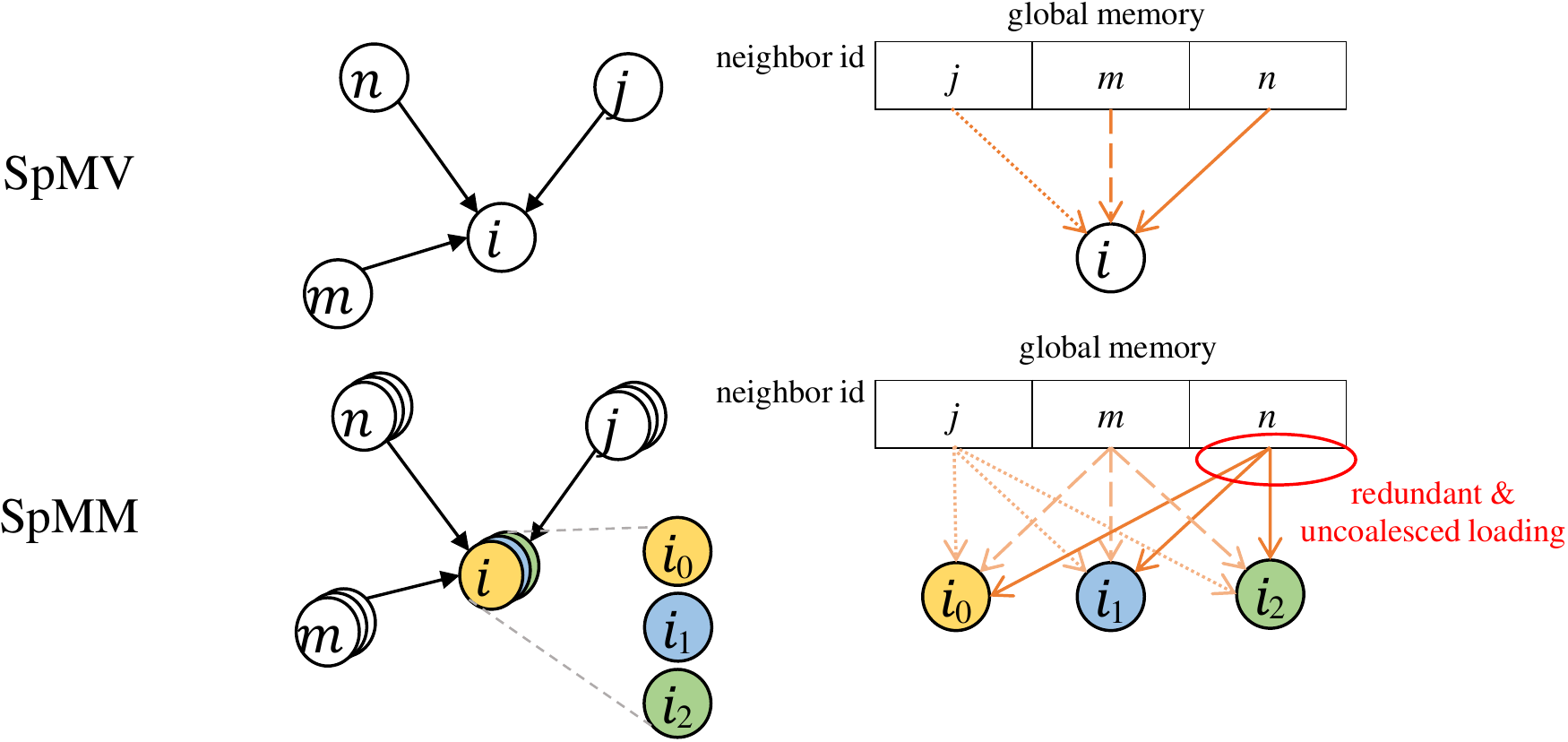}
\vspace{-5pt}
\caption{\modified{Differences in data loading in SpMV and its straightforward generalization to SpMM. SpMM exposes more redundant data loading and uncoalesced access pattern.}}
\vspace{-15pt}
\label{fig:coalesced-access}
\end{figure}

\begin{figure}[b]
\centering
\vspace{-15pt}
\includegraphics[width=0.45\textwidth]{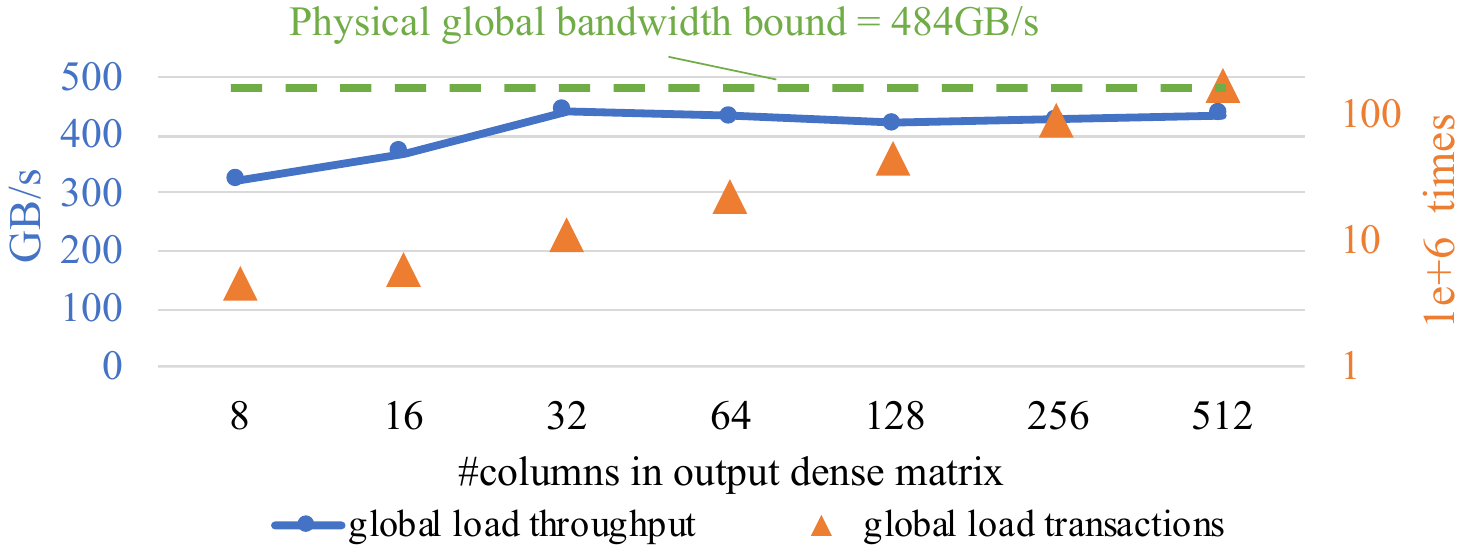}
\vspace{-5pt}
\caption{Profiling of \textit{csrmm2} in cuSPARSE. The loading throughput approaches upper bound when $N\geq32$ but memory transactions keep growing linearly.}
\vspace{0pt}
\label{fig:memory-load}
\end{figure}

From the GPU hardware perspective, SpMM exposes column-wise parallelism in output dense matrix, which does not exist in the widely-studied Sparse Matrix-Vector Product (SpMV). A straightforward generalization by adding parallel threads along the column dimension can result in uncoalesced access patterns to sparse matrix data, as shown in Fig.~\ref{fig:coalesced-access}. Uncoalesced access pattern has been proved to be inefficient on GPUs~\cite{cuda,kirk2016programming}. Thereby, SpMM kernel needs to be carefully designed to enable a coalesced access pattern for data loading. On the other hand, reusing sparse matrix data is crucial for SpMM, while this issue does not rise for SpMV. In real applications, the column dimension of the dense matrix can be up to 512~\cite{roc}, in which case the amount of memory transactions is substantial. Fig.~\ref{fig:memory-load} shows the profiling of SpMM kernel in cuSPARSE. We use as input sparse matrix a synthetic random matrix of 65K rows and 650K non-zeros, detailed in Section~\ref{sec:eval:B}, and test a range of column numbers (\modified{$N$} in Table~\ref{tb:notation}) in the dense matrix. The two metrics are 1) number of global load transactions (in the unit of 32bytes), and 2) global load throughput (in the unit of GB/s), both reported by Nvidia nvprof~\cite{cuda}. The GPU we use has a maximum global bandwidth of 484GB/s. Fig.~\ref{fig:memory-load} shows that the total number of memory transactions linearly grows with \modified{$N$}, but the kernel reaches near maximum bandwidth throughput after \modified{$N$} reaches 32. From the test we can observe that, unlike SpMV which is typically bounded by low bandwidth utilization~\cite{nvspmv}, SpMM can easily achieve a high utilization but suffers from too much data movement. Thereby, SpMM design requires a data-reuse mechanism to reduce redundant data transactions.

In this paper, we present GE-SpMM (an acronym for General-purpose SpMM), a customized CSR-based (Compressed Sparse Row) SpMM design that tackles all these challenges. We summarize our contributions as follows: 

\begin{itemize} 
    \item We present GE-SpMM, an efficient CSR-based SpMM-like kernel on GPUs to accelerate GNN workloads. GE-SpMM can be integrated into existing GNN frameworks with no data conversion overhead for various GNN algorithms.
    
    \item We introduce the Coalesced Row Caching (CRC) method for SpMM, 
    which uses GPU shared memory to cache sparse matrix rows. This method enables coalesced memory access to both sparse and dense matrix, leading to a more efficient utilization of bandwidth. The average improvement by adopting this method can be up to 1.25$\times$. 
    
    \item We introduce the Coarse-grained Warp Merging (CWM) method for SpMM, which reuses loaded sparse matrix by merging the workload of different warps. This technique reduces the amount of memory transactions and improves instruction-level parallelism. The average speedup by adopting this method can be up to 1.51$\times$.
    
    \item We conduct extensive experiments on GE-SpMM on real-world graphs \cite{snap,yang2016revisiting}. GE-SpMM achieves up to 1.41$\times$ speedup over Nvidia cuSPARSE~\cite{cusparse} and up to 1.81$\times$ over GraphBLAST~\cite{design}. We also embed GE-SpMM in GNN frameworks and get up to 3.67$\times$ CUDA time reduction on popular GNN models like GCN~\cite{gcn} and GraphSAGE~\cite{sage}.

\end{itemize}

The rest of this paper is organized as follows. Background information is introduced in Section~\ref{sec:bg}. The designs and optimizations of GE-SpMM will then be detailed in Section~\ref{sec:impl}, followed by the method to embed GE-SpMM in GNN frameworks. Our experimental setup and results will be presented in Section~\ref{sec:eval}. The paper is concluded in Section~\ref{sec:conc}.

\begin{table}[t]
\vspace{-10pt}
\caption{Notations.}\label{tb:notation}
\vspace{-10pt}
\begin{center}
    \begin{tabular}{ c |l}
    \hline
    Notation&Description\\
    \hline
    $A$&Sparse input matrix with dimension $M \times$ \modified{$K$}\\
    \hline
    $B$&Dense input matrix with dimension \modified{$K \times N$}\\
    \hline
    $C$&Dense output matrix with dimension $M \times$ \modified{$N$}\\
    \hline
    $M$&Number of rows in $A$. \\
    & Number of vertices in graph.\\
    \hline
    \modified{$K$}&Number of columns in $A$, \\
    & equal to $M$ in graph problems.\\
    \hline
    \modified{$N$}&Number of columns in $B$.\\
    & Feature vector length.\\
    \hline
    $nnz$&Number of non-zero elements in A.\\
         &Number of directed edges in graph.\\
    \hline
    \end{tabular}
\end{center}
\vspace{-20pt}
\end{table}
\section{Backgrounds and Related Works} \label{sec:bg}

In this section, we introduce background information about both SpMM and GNNs on GPUs. The notations used in this paper are shown in Table~\ref{tb:notation}.

\subsection{GPU Preliminaries}\label{sec:bg:gpu}

We use Nvidia GPUs with CUDA~\cite{cuda} as the programming interface in this paper. GPU is a highly parallel architecture composed of many streaming multiprocessors (SMs). An SM executes threads in a SIMT (Single Instruction Multiple Threads) fashion, and a bunch of 32 threads called a \textit{warp} run simultaneously. The \textit{warp} is transparent in CUDA programming model. Instead, users define a bunch of parallel \textit{blocks} and assign each \textit{block} with a certain number of threads. In CUDA programming model, each \textit{block} owns a \textit{shared memory}, which is more efficient to access than the global memory (accessible to all blocks). \textit{Shared memory} can be used for data reusing for different threads in order to reduce data transactions from the global memory.

Organizing threads into warps also have effects on the memory access pattern. GPU always try to merge the memory request from a warp into as few transactions as possible. From the program perspective, it is recommended in technical materials~\cite{cuda,kirk2016programming} to make a warp of threads access consecutive, aligned memory in one SIMT instruction. This technique is called \textit{coalesced memory access}.

\subsection{SpMM on GPUs} \label{sec:bg:spmm}

Because the dense matrix in the SpMM problem can be treated as a vector of vectors, a straight forward SpMM implementation is simply to perform SpMV for multiple times sequentially, as can be done in \cite{gunrock}. This method clearly does not exploit parallelism along the output column dimension. SpMV design in \cite{nvspmv} uses a GPU warp to process a row in the sparse matrix, and previous SpMM design in GraphBLAST~\cite{design} inherits this method so that multiple threads can work on one output row in parallel. 
For intra-warp data reuse, it uses a warp-level intrinsic (\_\_shfl) to broadcast fetched data to other threads within the same warp. However, GraphBLAST fails to consider reusing the sparse row data among different warps and still has room for improvement. Nvidia cuSPARSE library~\cite{cusparse} also provides a high-performance (not open-source) SpMM kernel (\textit{csrmm2}), but general SpMM-like operations in GNN applications are not supported in cuSPARSE. 

There are also other researches on high-performance SpMM kernels that perform better than cuSPARSE. \modified{Unlike GraphBLAST~\cite{design} which takes Compressed Sparse Row (CSR) format as input, these works require preprocess on input sparse matrix to form a new sparse format specially for SpMM, such as ELLPACK-R in Fastspmm~\cite{ortega2014fastspmm}, and sparse formats used by RS-SpMM~\cite{rsspmm} and ASpT~\cite{aspt}.} But they are not practical for GNN frameworks to adopt. These non-standard formats lead to extra memory space and difficulties in software maintenance. Moreover, preprocess time can be up to 5$\times$ actual SpMM computation time~\cite{rsspmm,aspt}. Although this cost can be tolerated in iterative algorithms, GNN applications sometimes demand running SpMM only a few times for one matrix. One example scenario is GNN inference, where trained models are directly used on new graphs to make predictions, such as predicting properties on new protein graphs~\cite{sage}. Another is sampled batch training~\cite{sage,fastgcn}, where the sampled subgraphs are different for each batch. For these applications, preprocess cannot be amortized in GNN frameworks since the benefit cannot make up to overhead.

\subsection{Graph Neural Network Frameworks} \label{sec:bg:framework}

Many existing systems aim to provide high performance and easy programming abstractions for GNN algorithm developers. Projects like DGL~\cite{dgl} and Pytorch-Geometric (PyG)~\cite{pyg} provide graph APIs on top of deep learning frameworks (e.g., Pytorch~\cite{pytorch}). Other systems in the industry \cite{aligraph, neugraph} and academia \cite{roc, gunrock} also provide programming interfaces and optimizations for GNNs. 

As SpMM is a critical operation in many GNN models, DGL and PyG both implement custom SpMM kernel instead of using sparse matrix operators provided by Pytorch. 
\begin{itemize}
    \item DGL internally calls the function, \textit{csrmm2} in cuSPARSE~\cite{cusparse}, to perform SpMM. However, for SpMM-like operations cuSPARSE does not provide corresponding functions, so DGL falls back to its own kernels. 
    \item Besides limited support for SpMM-like operations, \textit{csrmm2} produces a column-major output. Since in GNNs both input and output of feature matrix need to be row-major, DGL calls a matrix transpose from cuBLAS~\cite{cublas} to transform the layout.
    \item PyG uses another abstraction called MessagePassing to represent graph propagation in GNN models. Message-passing first generates message on all edges explicitly and then reduce them, while SpMM can fuse these two stages into one kernel. The consideration of MessagePassing is to allow more flexible user-defined operation, but with generality it loses the room for improving the performance of specific operations like SpMM. 
\end{itemize}


Some aforementioned researches on fast SpMM design require preprocessing~\cite{rsspmm,aspt}. CuSPARSE~\cite{cusparse} has limited support for SpMM-like operation~\cite{cusparse}. Other SpMM designs inherited from SpMV fail to consider the column-wise parallelism or sparse matrix data reuse, which may lead to inefficiency when implemented on GPUs. These will further hinder the performance of both SpMM and GNNs. 

\section{GE-SpMM Design} \label{sec:impl}

In Section~\ref{sec:intro}, we observe from experiments that SpMM can be memory-bounded, so it is crucial to load data in a more efficient way and reduce total memory transactions. In this section, we propose Coalesced Row Caching (CRC) and Coarse-grained Warp Merging (CWM) methods to achieve these goals. 

\begin{figure}[!t]
\centering
    \includegraphics[width=0.4\textwidth]{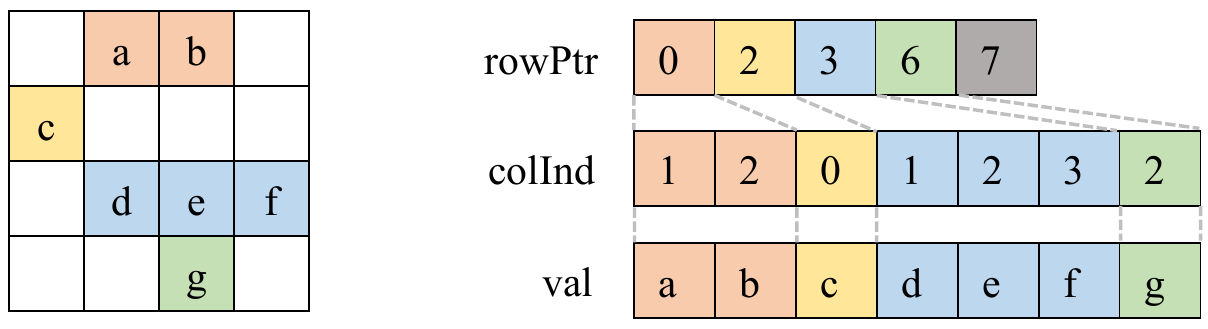}
    \vspace{-5pt}
    \caption{The sparse matrix (left) and its CSR representation (right).}
    \vspace{-15pt}
    \label{fig:csr}
\end{figure}

\begin{algorithm}[!b]
\caption{A simple parallel CSR-based SpMM}\label{alg:simple-spmm}
\begin{algorithmic}[1]
\renewcommand{\algorithmicrequire}{Input:}
\renewcommand{\algorithmicensure}{Output:}
\REQUIRE A.rowPtr[], A.colInd[], A.val[], B[]
\ENSURE C[]
\FOR {i = 0 \textbf{to} $M-1$ \textbf{in parallel}}
\FOR {j = 0 \textbf{to} $N-1$ \textbf{in parallel}}
\STATE result = 0
\FOR {ptr = A.rowPtr[i] \textbf{to} A.rowPtr[i+1]}
\STATE k = A.colInd[ptr]
\STATE result += A.val[ptr] * B[k,j]
\ENDFOR
\STATE C[i,j] = result
\ENDFOR
\ENDFOR
\end{algorithmic}
\end{algorithm}

\subsection{Data Organization in GE-SpMM}\label{sec:impl:csr}

In order to meet the requirement of compatibility to GNN frameworks with low preprocess overheads, an universal data format for both SpMM and other sparse matrix operations is required. The Compress Sparse Row (CSR) format is a widely-used format in vendor libraries (e.g., Nvidia cuSPARSE~\cite{cusparse}), data science toolkit (e.g., SciPy~\cite{scipy}), and GNN frameworks~\cite{dgl,gunrock}. As shown in \ref{fig:csr}, a sparse matrix is stored by three arrays using the CSR format: $rowPtr$, $colInd$, $val$. The column indices and values of non-zeros are first packed along the column and then stored in the order of their row indices. $rowPtr$ stores the offset of first element of each row in $colInd$. 

The data structure of CSR format determines the procedure of SpMM. The computation of each output $C[i,j]$, which is dot-product of sparse row $i$ of $A$ and dense column $j$ of $B$, begins with accessing $A.rowPtr$ for the offset of row $i$. Then the program needs to traverse a segment of $A.colInd$ and $A.val$ array to acquire the non-zeros in sparse row $i$. For each non-zero element, the program needs to use the column index got from $A.colInd$ to locate a specific row in dense matrix $B$. If the column index gives $k$, the program then loads $B[k,j]$, multiply it with the sparse element value from $A.val$ and add to final result. When all non-zeros in sparse row $i$ is consumed, the final result of $C[i,j]$ is returned. Algorithm \ref{alg:simple-spmm} shows this procedure with pseudo code.

\begin{algorithm}[!t]
\caption{\modified{SpMM with CRC}} \label{alg:caching}
\begin{algorithmic}[1]
\renewcommand{\algorithmicrequire}{\textbf{Input:}}
\renewcommand{\algorithmicensure}{\textbf{Output:}}
\REQUIRE A.rowPtr[], A.colInd[], A.val[], B[]
\ENSURE  C[]
\STATE i = tb\_id
\STATE j = tid
\STATE lane\_id = tid \% warp\_size
\STATE sm\_base = tid - lane\_id
\STATE row\_start = A.rowPtr[i]
\STATE row\_end = A.rowPtr[i+1]
\STATE result = 0
\FOR{ptr = row\_start \textbf{to} row\_end-1 \textbf{step} warp\_size }
\STATE /*load A.colInd and A.val with tile warp\_size*/
\IF{ptr+lane\_id$<$row\_end}
\STATE sm\_k[tid]=A.colInd[ptr+lane\_id]
\STATE sm\_v[tid]=A.val[ptr+lane\_id]
\ENDIF
\STATE \_\_syncwarp()
\STATE /*consume the loaded elements*/
\FOR {kk = 0 \textbf{to} warp\_size}
\IF{ptr+kk$<$row\_end}
\STATE k = sm\_k[sm\_base+kk]
\STATE result += sm\_v[sm\_base+kk] * B[k,j]
\ENDIF
\ENDFOR
\ENDFOR
\STATE C[i,j] = result
\end{algorithmic}
\end{algorithm}

\subsection{Coalesced Row Caching} \label{sec:impl:cache}

When trying to map Algorithms \ref{alg:simple-spmm} on parallel architectures like GPU, a simple way is to parallelize for-loop in line 1 and 2 since there exists no dependency among each iteration of the loop. For-loop at line 4 in Algorithm \ref{alg:simple-spmm} has variable loop bound decided at runtime and involves adding to one same variable, so it cannot be parallelized. As introduced in section \ref{sec:bg:gpu}, coalesced memory access can improve bandwidth efficiency, and it requires a warp of threads to access consecutive elements in one instruction. 

In Algorithm \ref{alg:simple-spmm}, it is easy to ensure coalesced access to dense matrix $B$ (the instruction at line 9). When we parallelize loop at line 1 and 2 among threads, we just need to ensure that threads within a warp have same $i$ and contiguous $j$. However, the current algorithm does not present coalesced memory access to any part of the sparse matrix. Since threads within a warp share same $i$, we cannot enable coalesced access to $A.rowPtr$. As to $A.colInd$ and $A.val$, in the current algorithm, 
the sequential execution of for-loop at line 4 forces threads in the same warp to access the same address 
(instructions at line 5-6), leading to the broadcast-like pattern in Fig~\ref{fig:coalesced-access}. 
Compared with ideal coalesced access to sparse rows (referring to a segment of $A.colInd$ and $A.val$), the current algorithm is not making full use of data in one memory transaction and issues too many transactions.

Our solution is to partially unroll this sequential for-loop, by a factor of $warp\_size$ (number of threads within a warp, 32 on Nvidia GPUs). During each iteration, a warp of threads first loads a tile of the sparse row $i$ into GPU shared memory. The size of a tile is the same as $warp\_size$, meaning that each thread loads a different element. For now, we assume that the total number of non-zeros in row $i$ is multiples of $warp\_size$, but we will generalize to arbitrary sparse row length later. After loading a tile to shared memory, threads enter an inner for-loop and compute on the loaded data one-by-one, but this time the sparse row data are loaded from shared memory instead of global memory. 

\begin{figure}[!t]
    \centering
    \includegraphics[width=0.48\textwidth]{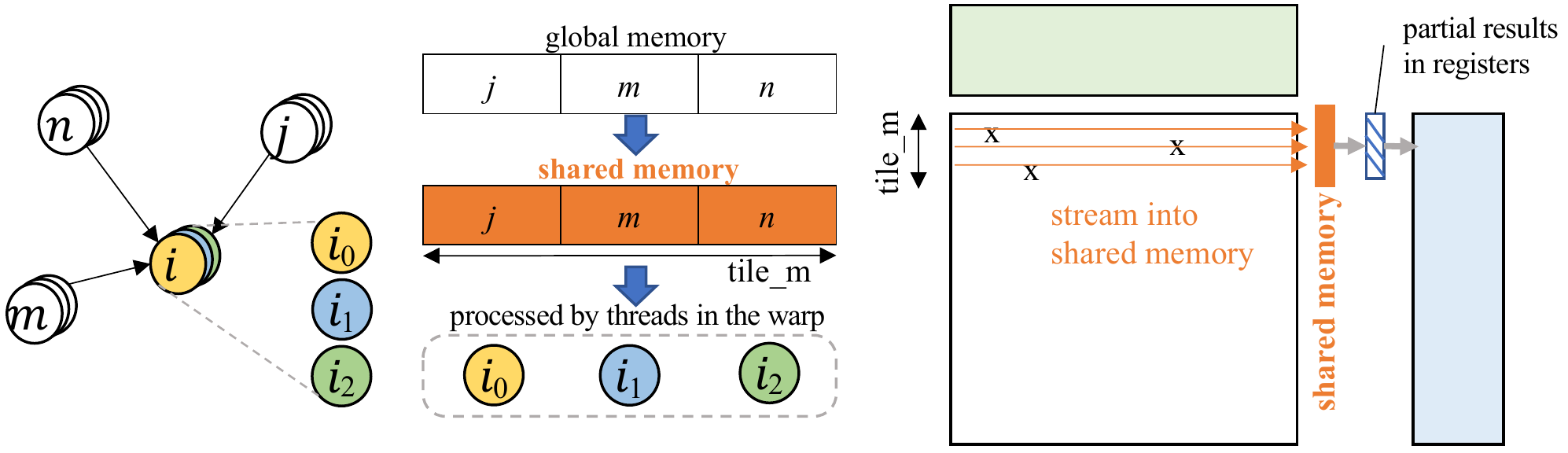}
    \vspace{-5pt}
    \caption{Overview of CRC which enables coalesced access to sparse rows. Non-zero elements of sparse matrix are first loaded into shared memory in a coalesced way and consumed later. Partial-sums are stored in local variables and written to global memory at the end. Note that no two threads write to the same output, so no atomic operation is needed.}
    \vspace{-15pt}
    \label{fig:cache}
\end{figure}

A pseudo-code of our method is listed in Algorithm \ref{alg:caching}. Basically, our method uses a two-phase strategy to load and compute on sparse rows, as shown in Fig.~\ref{fig:cache}. In the first phase, a warp loads a tile of a sparse row into shared memory to enable coalesced loading of a sparse row. In the second phase, a warp sequentially consumes the previously loaded elements. This is to ensure that a warp always accesses the same row in the dense matrix. When the number of non-zeros in sparse row exceeds a tile, the two-phase procedure continues until all non-zeros are consumed. It is not hard to deal with arbitrary row length. Each thread has a copy of row length, and always checks if the bound is exceeded before loading from a sparse row in the first phase (if-condition at line 10). In the second phase, since elements stored in shared memory are consumed sequentially, we simply check with loop bound in every iteration (condition at line 17).

The improvement of this algorithm over Algorithm \ref{alg:simple-spmm} lies in more efficient global memory loading of sparse rows. Ideally, the total number of load requests to these two arrays can be reduced by a factor of $warp\_size$, because a tile of sparse row is loaded in one coalesced request. In reality, sparse rows are often not perfectly aligned, leading to more than one transaction for a tile. Moreover, many sparse rows have fewer non-zeros than a tile, and the number of load transactions in the Algorithm \ref{alg:simple-spmm} equals to number of non-zeros. Thus the reduction in load transactions for these short rows is strictly less than $warp\_size$. Despite all these factors, the improved algorithm still achieves an obvious reduction of load transactions and improves the efficiency of global bandwidth, as shown in Section \ref{sec:eval:B}.

\subsection{Coarse-grained Warp Merging} \label{sec:impl:coarsen}

With CRC, we can convert the uncoalesced access to sparse matrix into a more efficient, coalesced way. From the data reuse perspective, the benefit of CRC is to share loaded sparse matrix via GPU shared memory. Although each non-zero element in the sparse matrix can be used to compute an entire row in the output, our CRC method only makes loaded elements shared by threads within the same warp. The consideration behind is to reduce synchronization overhead. To safely use the shared memory, synchronization needs to be called to avoid read-write races (between line 12 and line 18 in Algorithm \ref{alg:caching}). If we allow different warps to use the same piece of data in shared memory, we need to insert synchronization in the entire thread block which brings significant overhead. Since CUDA provides warp-level fine-grained synchronization, which is much less expensive then block-level synchronization, we only make loaded data to be shared within a warp. The drawback is that different warps still perform redundant loading of sparse rows. 

\begin{figure}[!t]
    \centering
    \includegraphics[width=0.42\textwidth]{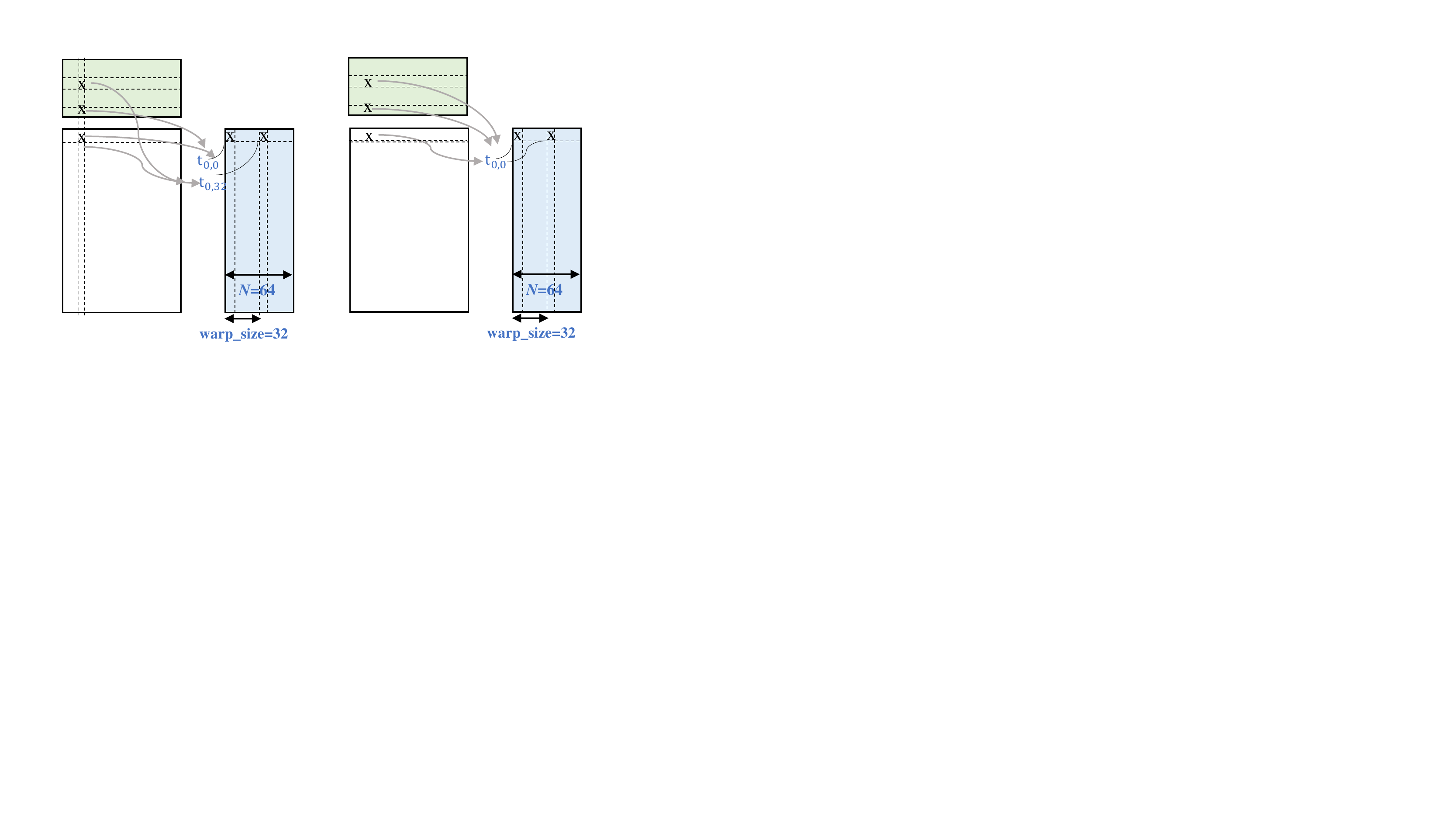}
    \vspace{-5pt}
    \caption{An example of CWM with $N=64$ and $CF=2$. Left: CWM is not applied, and the element in the sparse matrix is loaded twice by two threads. Right: the workloads of the previous two threads are merged, so the element in the sparse matrix is loaded only once.}
    \vspace{-15pt}
    \label{fig:coarsen}
\end{figure}

Our technique to address redundant data loading is Coarse-grained Warp Merging (CWM). It is similar to thread coarsening in dense matrix multiplication, which can improve bandwidth throughput with instruction-level parallelism (ILP)~\cite{volkov2008benchmarking}, and reduce the amount of memory transactions~\cite{yang2012unified}. The basic idea is to merge the workload of different warps that has redundant data loading. In SpMM, merging workloads means to make each thread produce more output. We illustrate CWM by an example in Fig. \ref{fig:coarsen}. In Fig. \ref{fig:coarsen}, the suffix of $t$ (short for thread) indicates both the indices to this thread and the indices of output it produces. Observe that $thread_{0,0}$ and $thread_{0,32}$ both load data from row 0 of $A$, but they belong to different warps and cannot share data under Algorithm \ref{alg:caching}. To dismiss this redundant load, we merge these two threads' workloads, making $thread_{0,0}$ compute both $C[0,0]$ and $C[0,32]$. $thread_{0,0}$ will have two partial sum variables locally, and with every non-zero element in $A$, it will load two values from matrix $B$ and update two partial results. In Algorithm \ref{alg:coarsen} we give a pseudo code with CWM adopted.

In Fig.~\ref{fig:coarsen}, we merge the workloads of two warps and cut the number of threads by half. Intuitively this process can continue and we can cut down more threads. We call the factor of thread number reduction \textit{coarsening factor} ($CF$). For example, $CF$ is 2 in Fig. \ref{fig:coarsen}, which means each thread is assigned to produce 2 output values. In general, the load transactions of sparse rows can be reduced by $CF$ through this technique. Another benefit of introducing thread coarsening is to improve bandwidth utility via ILP~\cite{volkov2008benchmarking}. Line 7-8 in Algorithm \ref{alg:coarsen} is independent memory loading instructions and GPU architecture can serve these two requests simultaneously, potentially increasing the usage of global bandwidth. Increasing $CF$ can further reduce memory load, but there will be fewer threads on the fly. Large $CF$ also causes each thread to hold more local variables for partial results, and this increment in resource usage may hurt performance. GPUs use massive parallel threads to hide all types of stalls, mostly the latency of memory load. When using CWM, it is significant to balance between data reuse and parallelism. Analytical models for choosing $CF$ could be difficult due to the entangled effects of hardware parameters and sparse matrix properties. We turn to an empirical method and experimented on our dataset of real-world graphs with $N=512$ to find a general best choice of $CF$, detailed in Section~\ref{sec:eval:B}.



\begin{algorithm}[!t]
\caption{SpMM with CRC and CWM($CF$=2)} \label{alg:coarsen}
\begin{algorithmic}[1]
\renewcommand{\algorithmicrequire}{\textbf{Input:}}
\renewcommand{\algorithmicensure}{\textbf{Output:}}
\REQUIRE A.rowPtr[], A.colInd[], A.val[], B[]
\ENSURE  C[]
\STATE /* initialization (line 1-6 of Algorithm 2) */
\STATE result\_1 = 0, result\_2=0
\FOR{ptr = row\_start \textbf{to} row\_end-1 \textbf{step} warp\_size}
\STATE /* load A.colInd, A.val (line 10-14 of Algorithm 2) */
\FOR {kk = 0 \textbf{to} warp\_size}
\STATE k = sm\_k[sm\_base+kk]
\STATE result\_1 += sm\_v[sm\_base+kk] * B[k,j]
\STATE result\_2 += sm\_v[sm\_base+kk] * B[k,j+warp\_size]
\ENDFOR
\ENDFOR
\STATE C[i,j] = result\_1
\STATE C[i,j+warp\_size] = result\_2
\end{algorithmic}
\end{algorithm}


\section{Accelerate GNN Frameworks with GE-SpMM}\label{sec:framework}

GE-SpMM is developed to accelerate GNN applications. The CSR format and SpMM-like operation support make it easy to be embedded in existing frameworks. This section briefly discusses how we use GE-SpMM to enhance the performance of existing GNN frameworks.

\subsection{GE-SpMM for Different Matrix Sizes}

\begin{figure}[!t]
    \centering
    \includegraphics[width=0.45\textwidth]{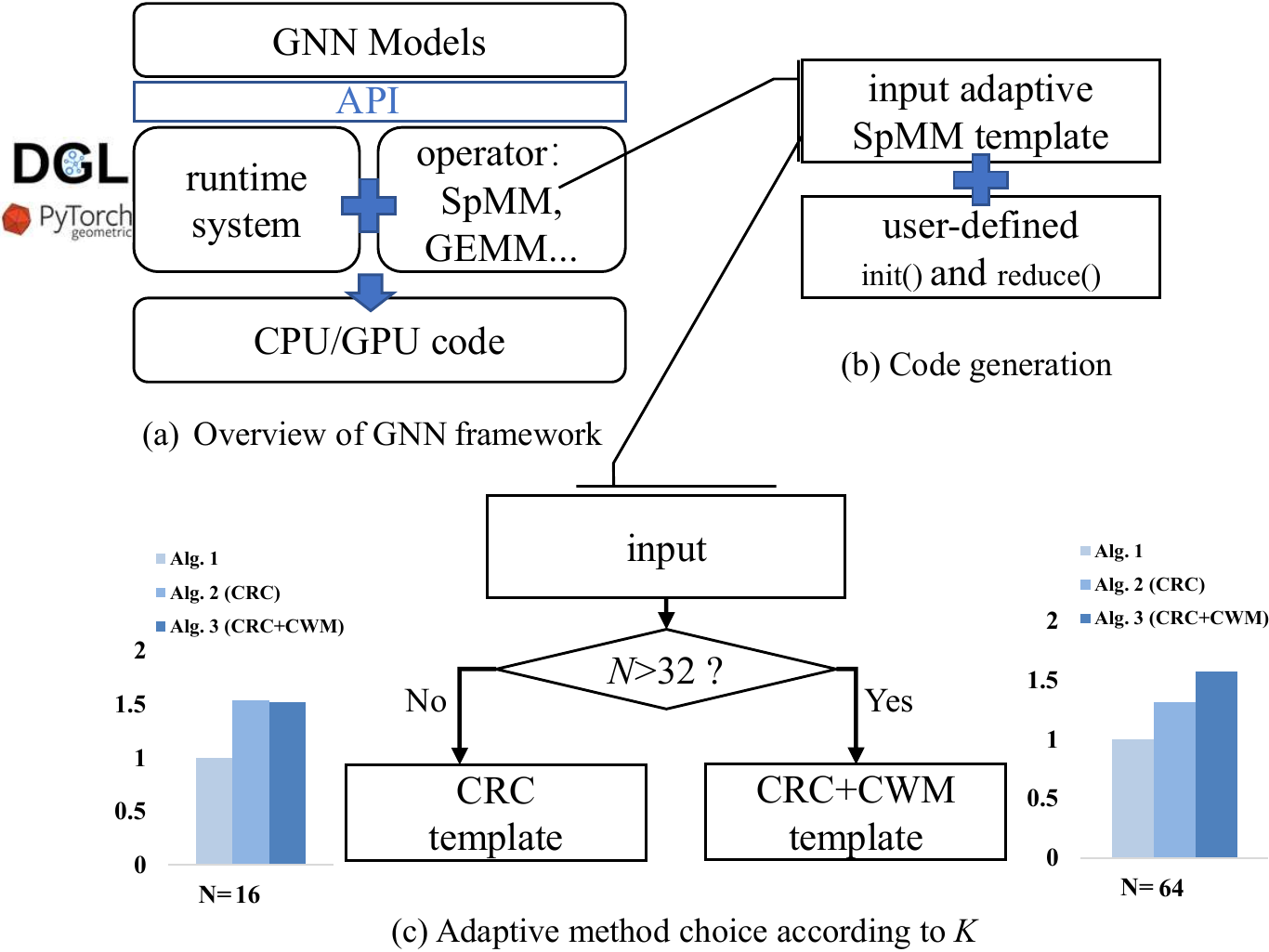}
    \vspace{-10pt}
    \caption{The overall flow of embedding GE-SpMM in GNN frameworks.}
    \vspace{-15pt}
    \label{fig:integration}
\end{figure}

To accelerate real applications, we make a few enhancement to support arbitrary input size ($N$). CRC and CWM apply well to problems with large $N$, where the kernel needs to load a large amount of data from dense matrix and is bottle-necked by bandwidth efficiency. \modified{Fig. \ref{fig:integration}(c) shows the overall benefit of our two techniques when $N=16$ and $N=64$, with average performance on the test dataset (detailed in Section~\ref{sec:eval:A}) normalized to Algorithm \ref{alg:simple-spmm}. When $N>32$, we apply both CRC and CWM in the kernel. CWM is not necessary for $N\leq 32$ since $warp\_size$ is 32, and we should directly call Algorithm \ref{alg:caching} to dismiss the overhead of unnecessary instructions.}

To address the need for SpMM-like operation in GNN models, we modify the basic GE-SpMM to allow user-defined operation. To define an SpMM-like operation, the user needs to provide an initialization function and a reduce function, both will be inlined at compile time. The parallel execution requires the reduction function to be associative and commutative, but common operations like taking sum or maximum are naturally valid. 

\subsection{GNN Acceleration Based on GE-SPMM}

Current GNN frameworks like DGL~\cite{dgl} and PyG~\cite{pyg} are often based on other deep learning (DL) frameworks (e.g. PyTorch~\cite{pytorch}), but add new APIs for graph operations. Although it is possible to express graph operations with sparse tensor operations provided by DL frameworks, the performance of sparse tensor operations are not satisfactory, so DGL implements all graph-related operations in C++/CUDA and exposes to DL frameworks as shared lib. We also follow this method to accelerate GNN application with a high-performance CUDA kernel. 

To be more specific, we wrap our kernel inside a custom autograd function, which is an atomic operator with gradient definition in PyTorch. This function represents an aggregation step on the graph, and can be used to build GNN layers and modules. Since DGL already implements SpMM-like in CUDA, we simply substitute their kernel with ours and rebuild the project. PyG is another popular Python library for GNN built on PyTorch. It abstracts GNN as MessagePassing procedure and implements MessagePassing as a versatile module that allows user-defined message and aggregation functions. MessagePassing is a more general interface than SpMM-like, so we cannot use SpMM-like to replace MessagePassing. We instead implement an SpMM-like operator and replace the MessagePassing function calls in training code with ours. An overview of how we integrate GE-SpMM to GNN frameworks is shown in Fig.~\ref{fig:integration}.

\section{Experiment Evaluations} \label{sec:eval}

We conduct extensive experiments on our GE-SpMM design, and the performance comparison against various SpMM kernels and GNN frameworks is shown in this section.

\subsection{Experiment Setup} \label{sec:eval:A}

\subsubsection{Graph Benchmarks}

In order to test the proposed GE-SpMM for GNN workloads, we run experiments on three graphs~\cite{yang2016revisiting} used for node-classification tasks in many GNN models~\cite{gcn, sage}, \verb|Cora|, \verb|Citeseer|, and \verb|Pubmed|, whose properties are listed in Table \ref{tb:gnn-graph}. We also test the performance on SNAP dataset collected in SuiteSparse Matrix Collection~\cite{suitesparse}, a sparse matrix benchmark. SNAP group in SuiteSparse contains 66 valid graphs from various domains. The original SNAP dataset maintained in \cite{snap} also contains metadata of some graphs which are not collected in SuiteSparse, but we limit our test to graphs in SuiteSparse to save the effort of converting metadata to standard input. We omit two graphs (\verb|FriendSter| and \verb|Twitter|) due to out-of-memory. 
Some items in SuiteSparse contains more than one matrices. We only run tests on one matrix in each item\footnote{The matrix which has the same filename as this item is considered the default one.}. This set of 64 sparse matrices has size $M$ from 1005 to 4847571 with $nnz/row$ from 1.58 to 32.53. \cite{roc} experiments on models with feature size up to 512 and presents the best model accuracy of feature size around 256 with deeply-stacked layers. This should provide an intuition of how large $N$ is in real applications. \modified{All average results are based on the geometric mean.}

\begin{table}[h]
\vspace{-10pt}
\caption{Graphs Used in GNN for Classification~\cite{yang2016revisiting}}
\vspace{-5pt}
\label{tb:gnn-graph}
    \centering
    \begin{tabular}{ c | c  c  c}
        \hline
         Graph & \# Vertices & \# Edges & \# Classes \\
         \hline
         \verb|Cora|       & 2708 & 5429 & 7 \\ 
         \verb|Citeseer|   & 3327 & 4732 & 6 \\
         \verb|Pubmed|     & 19717 & 44338 & 3 \\
         \hline
    \end{tabular}
\vspace{-5pt}
\end{table}

\subsubsection{SpMM Baselines}

We compare our GE-SpMM with the following baselines in our experiments.

\begin{itemize}
    \item \textbf{SpMM kernel by vendor:} \textit{csrmm2}. It is a function in cuSPARSE~\cite{cusparse} for SpMM. cuSPARSE has two functions for multiplication of sparse and dense matrices. The \textit{csrmm2} assumes a row-major input dense matrix, while the other one, \textit{csrmm}, assumes a column-major input dense matrix. The \textit{csrmm2} consistently outperforms \textit{csrmm}, and here we show a comparison to \textit{csrmm2}. Note that, the output dense matrix of \textit{csrmm2} is column-major, which is a convention in many Nvidia libraries. As explained in Section~\ref{sec:bg:framework}, GNN applications require row-major output, so existing solution is forced to perform matrix transpose upon \textit{csrmm2} output. We do not add this to baseline when comparing kernel performance. However, this overhead of cuSPARSE cannot be ignored in real applications.
    
    \item \textbf{Open-source SpMM kernel:} \textit{rowsplit} in GraphBLAST~\cite{design}. It is a most-recent CSR-based SpMM implementation in literature. 
    
    \item \textbf{Graph processing engine on GPUs:} GunRock~\cite{gunrock}. Because the SpMM can be executed from a graph perspective by assigning each vertex with a feature vector, we also compare our GE-SpMM with the state-of-the-art graph processing system on GPUs. 
    
\end{itemize}

\subsubsection{Environments}

We conduct experiments on the following two machines:  
\begin{itemize}
	\item Machine 1. GPU: Nvidia GTX 1080Ti, Compute Capability 6.1 (28 Pascal SMs at 1.481 GHz, 11 GB GDDR5X with 484 GB/s bandwidth). Host CPU: Intel(R) Xeon(R) CPU E5-2643 v4 (24 cores).
	\item Machine 2. GPU: Nvidia RTX 2080, Compute Capability 7.5 (46 Turing SMs at 1.515 GHz, 8 GB GDDR6 with 448 GB/s bandwidth). Host CPU: Intel(R) Core(TM) i7-9700K (8 cores).
\end{itemize}

In kernel performance tests, all codes are compiled using NVCC (CUDA compiler provided by Nvidia) in CUDA 10.1 with \verb|-O3| flag. Execution time is measured from average of 200 kernel runs. Throughput is calculated from theoretical float-operation ($2*nnz*N$) over measured execution time. For application speedup, all reported items (kernel and model time) refer to CUDA time reported in PyTorch profiler.

\subsection{Benefits of GE-SpMM Design} \label{sec:eval:B}
\subsubsection{Benefits of Coalesced Row Caching}

The aim of CRC is to enable coalesced access to sparse matrix and improve bandwidth efficiency. To evaluate the effectiveness of CRC, we use Nvidia's nvprof to profile two metrics~\cite{cuda}: gld\_transactions (GLT), the number of global load transactions; gld\_efficiency (GLT\_effi), ratio of requested global memory load throughput to required global memory load throughput. One issued memory transaction loads a fixed size of data, but the program may only require a part of it. This metric can reflect how efficiently the program uses global bandwidth. 

The tests run on three synthetic random graphs\footnote{The code to generate random graph is from repo of Ligra~\cite{shun2013ligra}.} with $N=512$. We only have profiling results on Machine 1 since nvprof in CUDA 10.1 does not support GPU over 7.2 capability~\cite{cuda}. The results in Table \ref{tb:eval:CRC} show that using CRC can significantly reduce the total number of load transactions, as well as improve the memory load efficiency due to coalesced memory access. 

\begin{table}[h]
	\caption{Effects of CRC}
	\label{tb:eval:CRC}
	\centering
	\begin{tabular}{c|c|c|c}
		\hline
		Matrix&Method& GLT($\times$32bytes)&GLT\_effi\\
		\hline
		$M$=16K   &w/o CRC& 1.34e+8 & 68.95\%\\
		\cline{2-4}
		$nnz$=160K&w/ CRC& 0.55e+8 & 92.40\%\\
		\hline
		M=65K&w/o CRC& 5.36e+8 & 68.95\%\\
		\cline{2-4}
		nnz=650K&w/ CRC& 2.18e+8 & 92.40\%\\
		\hline
		M=262K&w/o CRC& 21.47e+8 & 68.95\%\\
		\cline{2-4}
		nnz=2.6M&w/ CRC& 8.73e+8 & 92.39\%\\
		\hline
	\end{tabular}
\end{table} 

\begin{figure}[!t]
    \centering
    \vspace{-5pt}
    \includegraphics[width=0.48\textwidth]{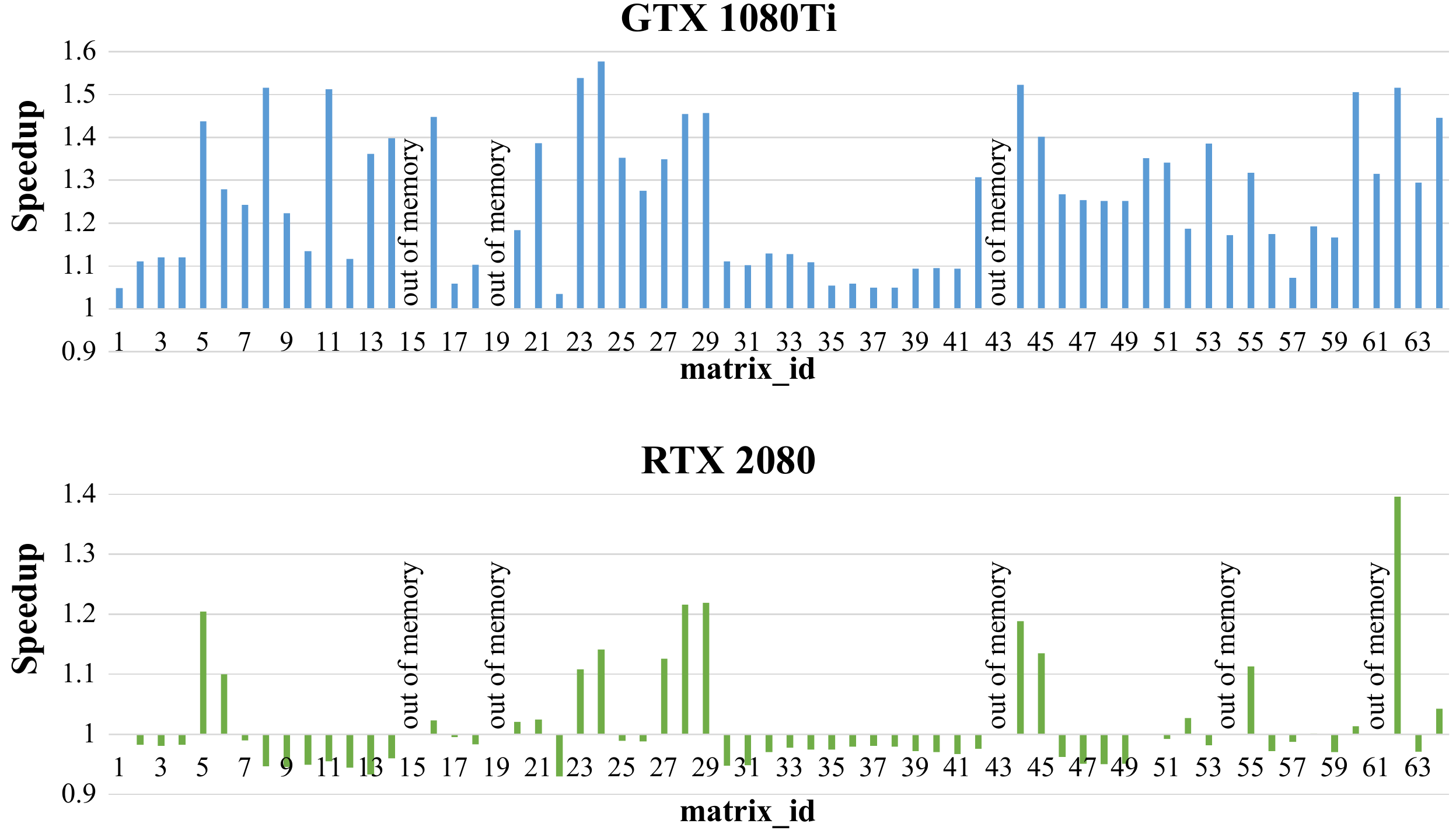}
    \vspace{-5pt}
    \caption{Relative speedup using Coalesced Row Caching.}
    \vspace{-15pt}
    \label{fig:eval:CRC}
\end{figure}

Fig.~\ref{fig:eval:CRC} shows the relative improvement of applying CRC (Algorithm \ref{alg:caching} against Algorithm \ref{alg:simple-spmm}). On GTX 1080Ti, CRC brings an average of 1.246$\times$ performance gain. On RTX 2080, simply applying CRC does not bring significant performance gain (average of 1.011$\times$). 
However, CRC is the foundation of our second technique CWM. One benefit of CWM is a high throughput of loading dense matrix with ILP, but without CRC, the bandwidth is spent on a large amount of inefficient access to the sparse matrix, and leaves little room for loading the dense matrix. In the next part, we will show that combined with CWM, GE-SpMM still achieved significant improvement over simple SpMM on RTX 2080.

\subsubsection{Benefits of Coarse-grained Warp Merging}

CWM is introduced in order to reuse loaded data and reduce global transactions. However, it reduces the total number of warps and may harm parallelism. We also use nvprof to test the effects brought by CWM. We profile three metrics: gld\_transactions (GLT), as previously mentioned it indicates the amount of data loading; gld\_throughput, the throughput of the global load; achieved\_occupancy (Occ), the ratio of the average active warps to the maximum supported on a multiprocessor, which can be a reflection of achieved parallelism. 

The results in Table \ref{tb:eval:CWM} are tested on one of the random graphs ($M=65K, nnz=650K, N=512$). It shows that CWM can reduce global data loading. When the coarsening factor ($CF$) gets larger, gld\_transactions keep decreasing, but the benefit becomes smaller because loading dense matrix takes up most of the transactions. When increasing $CF$, the occupancy also decreases, indicating parallelism loss. \modified{The combination of CRC and CWM brings an average of 1.65$\times$ and 1.53$\times$ speedup on GTX 1080Ti and RTX 2080 respectively over non-optimized version (Algorithm \ref{alg:simple-spmm}).}

\begin{figure}[!t]
    \centering
    \vspace{-5pt}
    \includegraphics[width=0.48\textwidth]{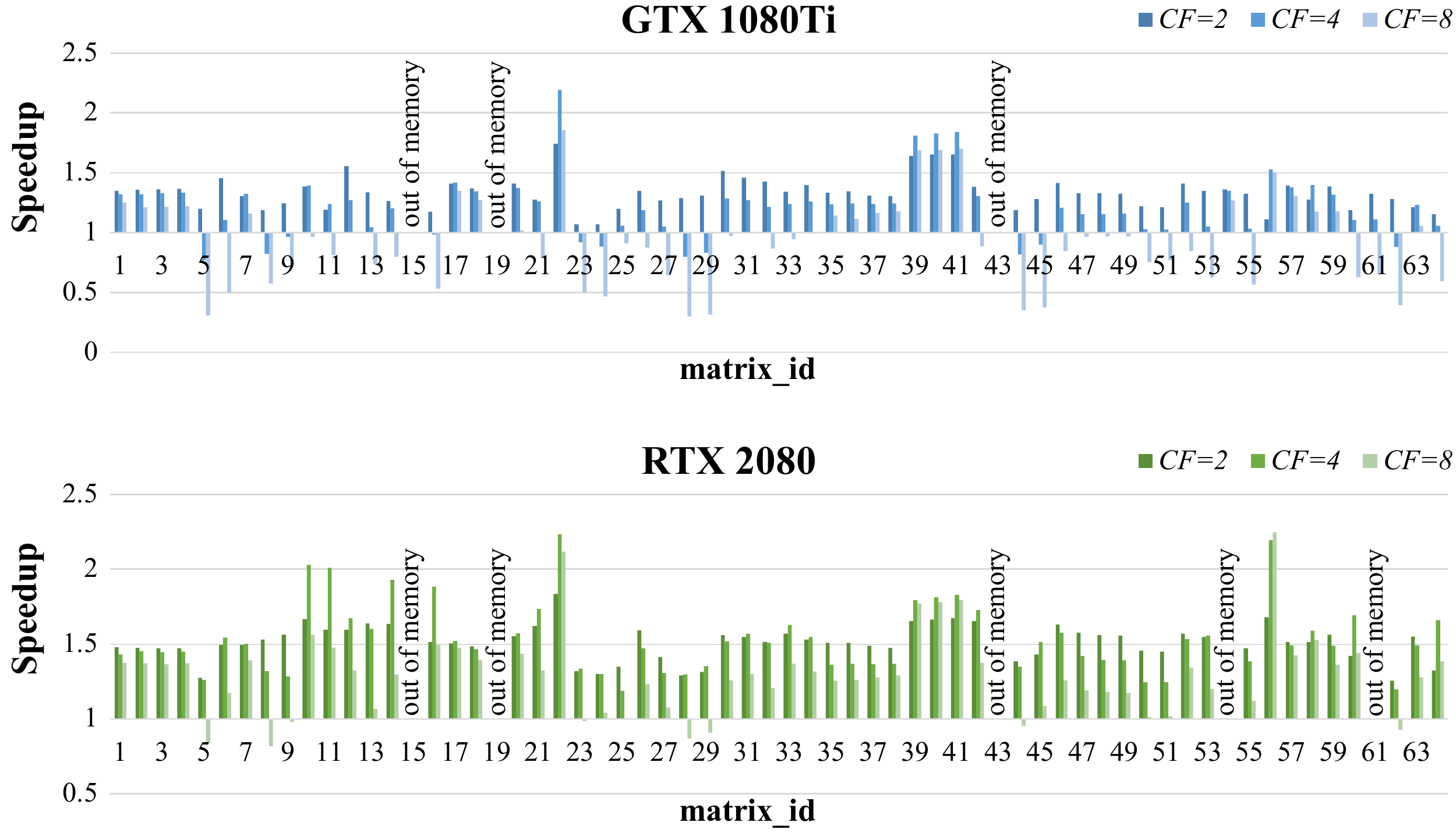}
    \vspace{-5pt}
    \caption{Relative speedup using Coarse-grained Warp Merging when $CF$ varies.}
    \vspace{-15pt}
    \label{fig:eval:CWM}
\end{figure}

\begin{table}[h]
\vspace{-10pt}
	\caption{Effects of CWM}
	\vspace{-5pt}
	\label{tb:eval:CWM}
	\centering
	\begin{tabular}{c | c | c | c}
		\hline
		Method&GLT($\times$32bytes)&gld\_throughput(GB/s) & Occ\\
		\hline
		w/o CWM & 2.18e+8 & 479.54 & 0.78\\
		\hline
		CWM (CF=2) & 1.93e+8 & 567.82 & 0.78\\
		\hline
		CWM (CF=4) & 1.80e+8 & 479.23 & 0.75\\
		\hline
		CWM (CF=8) & 1.74e+8 & 395.22 & 0.75\\
		\hline
	\end{tabular}
\vspace{-5pt}
\end{table}

\begin{figure}[b]
    \centering
    \vspace{-15pt}
    \includegraphics[width=0.5\textwidth]{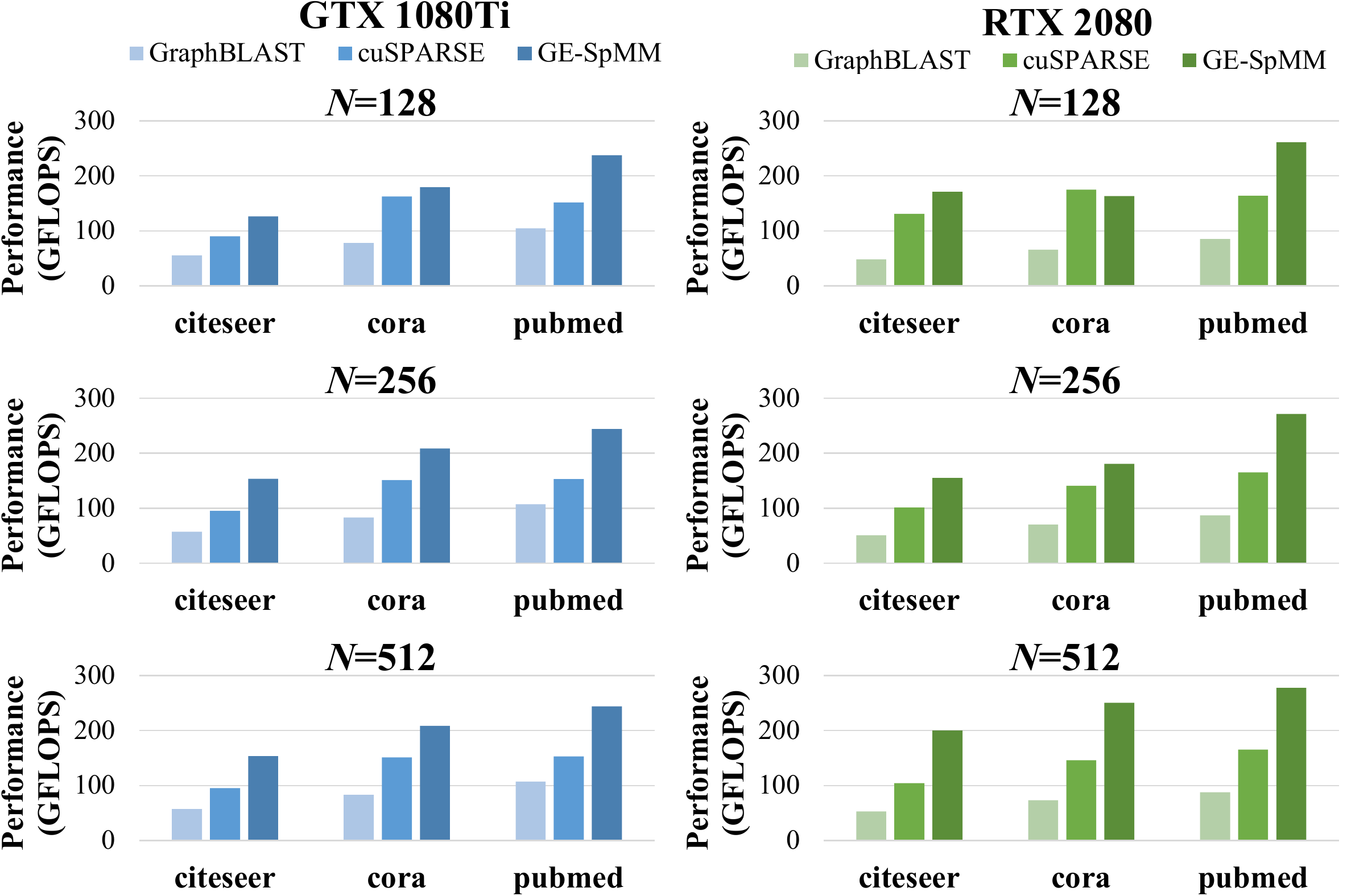}
    \vspace{-5pt}
    \caption{Performance on GNN graphs.}
    \vspace{-5pt}
    \label{fig:overall-gnn}
\end{figure}

\begin{figure*}[!t]
    \centering
    \vspace{-5pt}
    \includegraphics[width=0.98\textwidth]{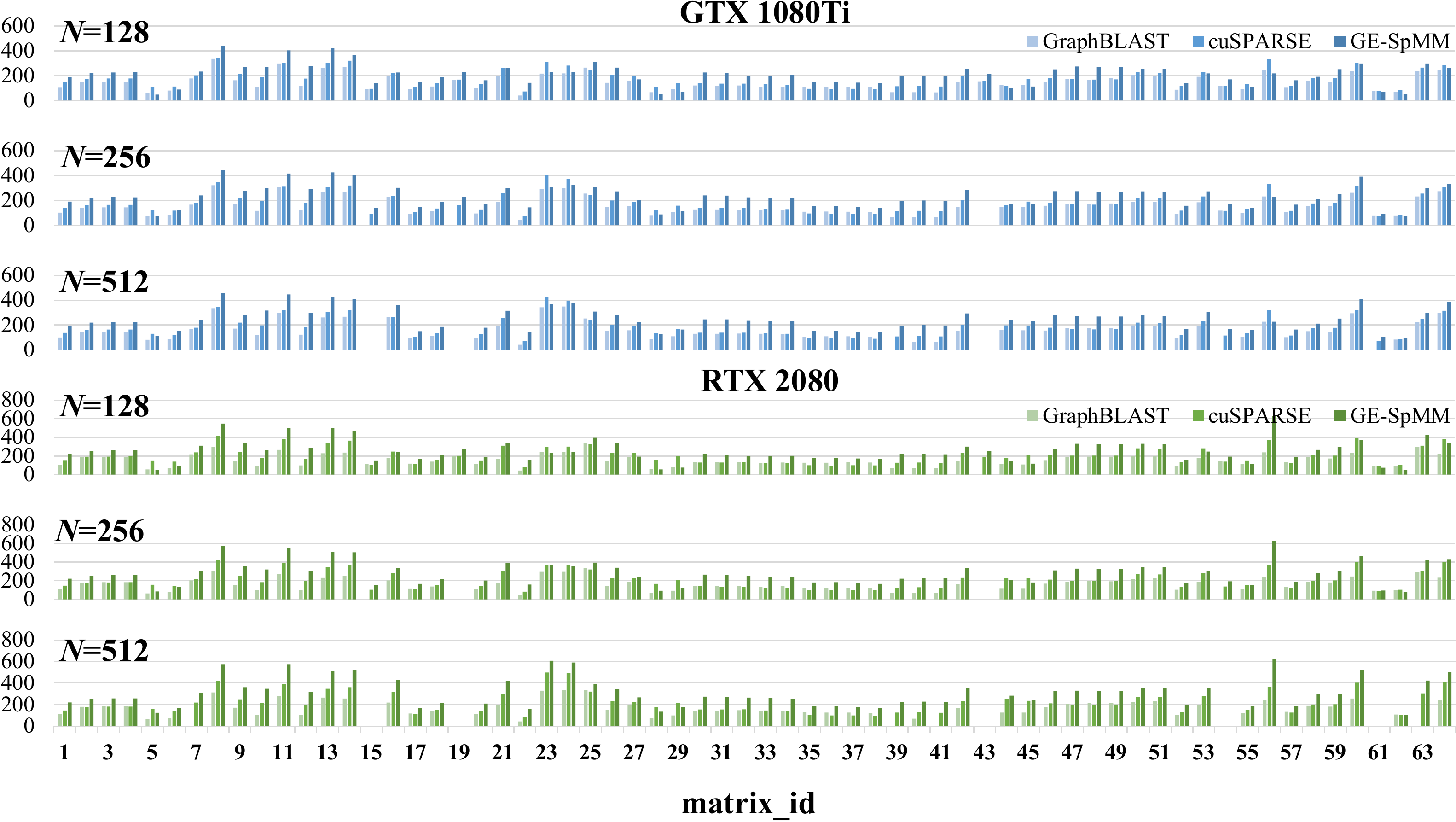}
    \vspace{-5pt}
    \caption{Overall performance on 64 graphs in SNAP dataset. The omitted bars are due to out-of-memory. The matrix\_id corresponds to the alphabetical order of matrix name. }
    \vspace{-5pt}
    \label{fig:overall-snap}
\end{figure*}

In practice, tuning for optimal $CF$ requires a balance between data reuse and parallelism, which are related to properties of the input matrix. In Fig.~\ref{fig:eval:CWM}, we plot the relative performance when taking different $CF$. Each bar represents a test on one graph in SNAP dataset. The relative performance means speedup over not using CWM. It can be observed that $CF=2$ works well for most matrices, while $CF>4$ shows obvious performance drop. For rare cases (4 and 1 out of 64 on two GPUs), choosing $CF=2$ causes over 15\% performance loss compared to optimal $CF$. Since our goal is to provide a runtime SpMM kernel, we avoid any parameter tuning and use $CF=2$ because it provides the best overall effect. The kernel performance reported in the following experiments is all the results of $CF=2$. Moreover, our previous tests show that on RTX 2080, CRC cannot bring large performance gain, but a combination of CRC and CWM can bring significant improvement (an average of 1.51$\times$) with $CF=2$.

\subsection{Overall Performance of SpMM Kernel} \label{sec:eval:C}

In this part, we present overall performance compared to kernels in cuSPARSE~\cite{cusparse} and GraphBLAST~\cite{design}. All matrices in tests are single-precision, as is the case in GNN models. For kernel performance, we test $N$ from 128 up to 512.

\subsubsection{Graphs for GNNs}

We test GE-SpMM on three graphs used in GNN models, and results are shown in Fig.~\ref{fig:overall-gnn}. 
Benefit from our two techniques, GE-SpMM can outperform cuSPARSE by at most 1.62$\times$. Tests on these three graphs show that GE-SpMM can potentially accelerate real GNN models if applied in GNN frameworks.

\subsubsection{Graphs from SNAP}

The performance comparison of GE-SpMM and baselines on SNAP dataset is shown in Fig.~\ref{fig:overall-snap}. To summarize, the average performance gain is listed in Table \ref{tb:average}, GE-SpMM achieves up to 1.43$\times$ and 1.81$\times$ speedup compared with cuSPARSE~\cite{cusparse} and GraphBLAST~\cite{design}. 
Although the arbitrary performance is related to graph characteristics, Fig.~\ref{fig:overall-snap} can demonstrate that for any specific graph, our kernel becomes more competitive over baseline when $N$ gets larger. In other words, our techniques are crucial for applications with large $N$.

\begin{table}[h]
    \centering
    \vspace{-5pt}
    \caption{GE-SpMM average improvement on SNAP dataset}
    \vspace{-5pt}
    \begin{tabular}{c|c|c|c|c}
        \hline
        \textbf{Machine} & \textbf{Baseline} & N=128 & N=256 & N=512\\
        \hline
        \multirow{2}{*}{GTX 1080Ti} & cuSPARSE~\cite{cusparse} & 1.18 & 1.30 & 1.37\\
        & GraphBLAST~\cite{design} & 1.42 & 1.44 & 1.61\\
        \hline
        \multirow{2}{*}{RTX 2080} & cuSPARSE~\cite{cusparse} & 1.20 & 1.34 & 1.43\\
        & GraphBLAST~\cite{design} & 1.57 & 1.73 & 1.81\\
        \hline
    \end{tabular}
    \vspace{-5pt}
    \label{tb:average}
\end{table}

\subsection{Comparison with Graph Engines on GPUs} 

\begin{figure}[h]
    \centering
    \vspace{-20pt}
    \includegraphics[width=0.48\textwidth]{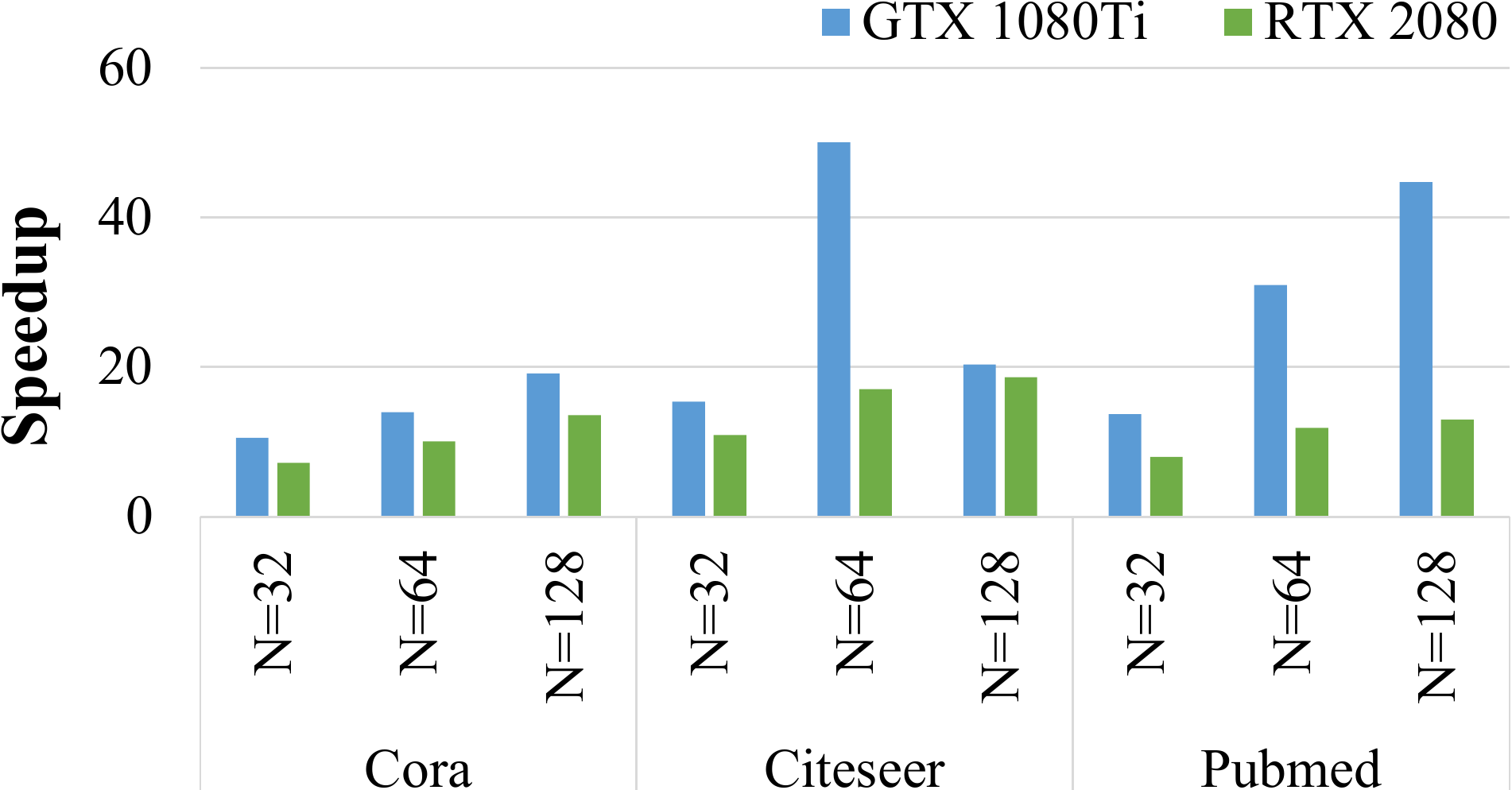}
    \vspace{-5pt}
    \caption{Speedup of GE-SpMM over GunRock-based SpMM.}
    
    \label{fig:gunrock}
\end{figure}

It is also possible to build SpMM with graph engines from a graph processing perspective. GunRock\cite{gunrock} provides many APIs and built-in kernels that allow users to write graph algorithms. We use GunRock's API \verb|advance| to write an SpMM program. However, GunRock does not provide any options to parallelize along feature dimension, because in traditional graph algorithms like PageRank, the feature of a vertex is an undividable scalar. GunRock fails to provide feature-dimension parallelism, which significantly harms SpMM performance. Fig.~\ref{fig:gunrock} shows the kernel execution time of GE-SpMM compared to the program written with GunRock. GE-SpMM outperforms GunRock-based implementation by 18.27$\times$ on average for all test cases, indicating that SpMM and GNN workloads require new primitives in graph processing frameworks rather than SpMV.

\modified{

\subsection{Comparison with preprocess-based approaches}

Some previous researches propose specialized sparse formats for SpMM problem that exploit regular memory access~\cite{ortega2014fastspmm} or data reuse~\cite{rsspmm,aspt}. As far as we know, ASpT~\cite{aspt} is the best SpMM implementation publicly available. ASpT~\cite{aspt} exploits reusing dense matrix data with tiling. It requires a special sparse format composed of CSR and additive arrays to mark the positions of locally-dense blocks explicitly. Our optimizations in this paper to reuse sparse matrix data is orthogonal to their techniques. 

We test their source code on our machines on the SNAP dataset and results are listed in \ref{tb:aspt}. Our GE-SpMM achieves average of 0.93X, 0.97X, 1.00X for N=128, 256, 512 on GTX1080Ti GPU, and average 0.85X, 0.93X, 0.98X on RTX2080, against ASpT (without preprocess time).

The preprocess overhead varies significantly on different sparse matrices, from 0.01$\times$ to 64.53$\times$ of actual SpMM execution time, and average overhead is 0.47$\times$ execution time on GTX1080Ti and 0.34$\times$ on RTX2080. With preprocess time added (one preprocess + one run), our kernel is average 1.43$\times$$\sim$2.06$\times$ against ASpT.

\begin{table}[!h]
    \centering
    \vspace{-5pt}
    \caption{\modified{GE-SpMM average speed against ASpT~\cite{aspt}}}
    \vspace{-5pt}
    \begin{tabular}{c|c|c|c|c}
        \hline
        \textbf{Machine} & \textbf{Baseline} & N=128 & N=256 & N=512\\
        \hline
        \multirow{2}{*}{GTX 1080Ti} & ASpT~\cite{cusparse} & 0.93 & 0.97 & 1.00\\
        & ASpT w/ preproc~\cite{design} & 1.88 & 1.97 & 2.06\\
        \hline
        \multirow{2}{*}{RTX 2080} & ASpT~\cite{cusparse} & 0.85 & 0.93 & 0.98\\
        & ASpT w/ preproc~\cite{design} & 1.43 & 1.57 & 1.69\\
        \hline
    \end{tabular}
    \vspace{-5pt}
    \label{tb:aspt}
\end{table}

}

\subsection{End-to-End Performance for GNNs} \label{sec:eval:E}

Our kernel is developed to accelerate GNN applications, and the CSR-based property makes it easy to be embedded in existing frameworks. As introduced in Section~\ref{sec:framework}, we embed GE-SpMM in DGL~\cite{dgl} and PyG~\cite{pyg}. We test application speedup on GCN, GraphSAGE, and other popular GNN models that involve SpMM or SpMM-like operations. 

\subsubsection{GNNs based on SpMM Operators}

\begin{figure}[t]
	\centering 
	\vspace{-5pt}
	\includegraphics[width=0.45\textwidth]{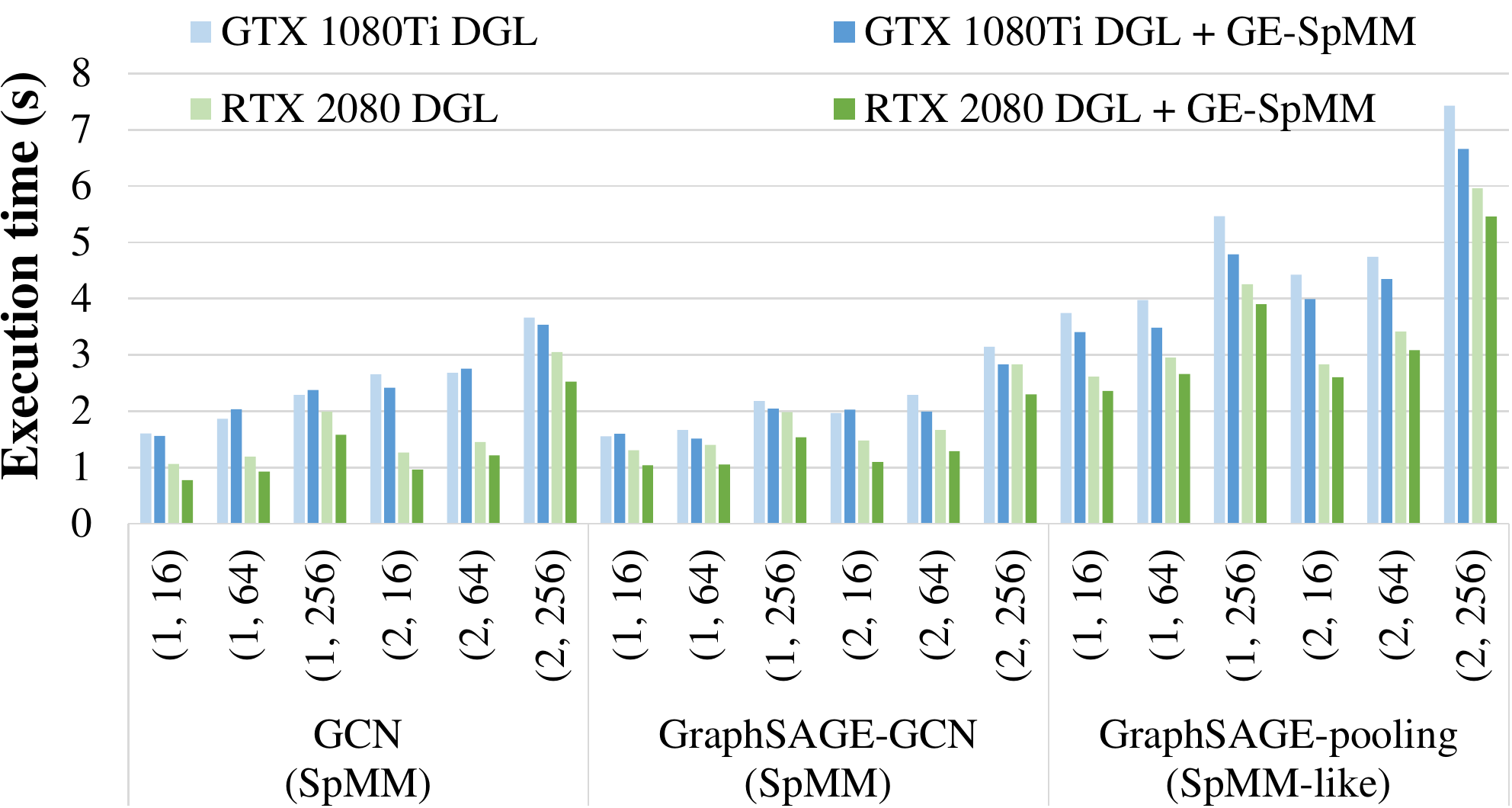}
	\vspace{-5pt}
	\caption{Accelerating GNNs using GE-SpMM in DGL~\cite{dgl}, (x, y) represents the number of layers (x) and the length of features (y) in a GNN model, which is the input parameter for GNNs.}
	\vspace{-10pt}
	\label{fig:eval:dgl}
\end{figure}

\textbf{Comparison with DGL.} SpMM is a very common operator in GNN, and GE-SpMM can benefit many GNN models. We test the benefit of integrating GE-SpMM in DGL with two models: GCN~\cite{gcn} and GraphSAGE-gcn\cite{sage}. For both models, we use example codes provided by DGL, and run tests on different model settings (number of layers and length of feature vectors in each layer). We use PyTorch profiler to monitor the entire training process and record total CUDA time from the report. The results are shown in Fig.~\ref{fig:eval:dgl}. GE-SpMM brings speedup in most of these applications. However, in 4 tests on GTX 1080Ti, GE-SpMM does not bring acceleration over original DGL. This is because the feature length of the last layer in GNN models usually equals the number of classes in the classification problem. Therefore, the output layer involves SpMM with a small $N$, when GE-SpMM is not very competitive.

\begin{figure}[t]
	\centering 
	\vspace{-5pt}
	\includegraphics[width=0.45\textwidth]{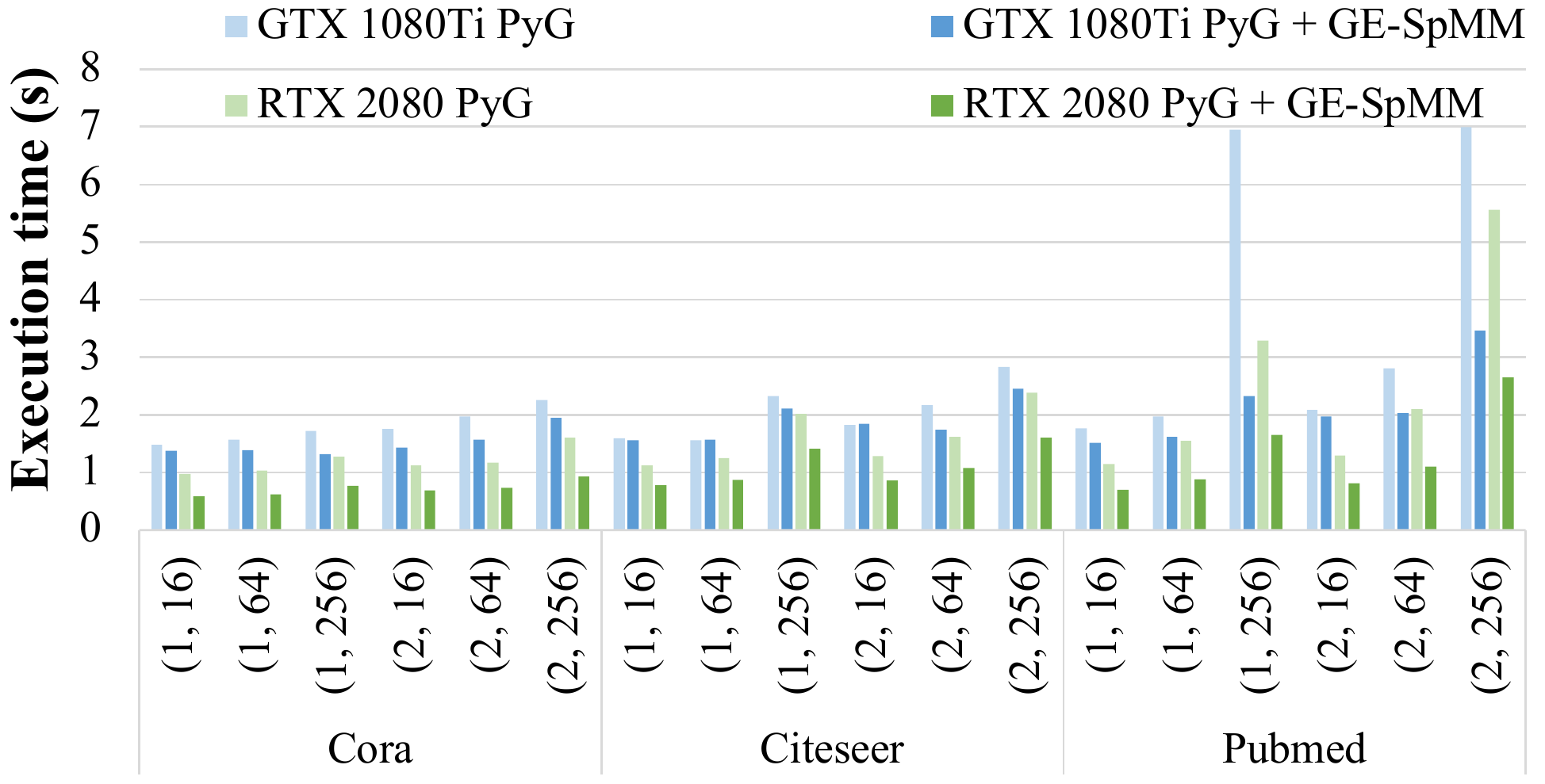}
	\vspace{-5pt}
	\caption{Accelerating GNNs using GE-SpMM in PyG~\cite{pyg}, (x, y) represents the number of layers (x) and the length of features (y) in a GNN model, which is the input parameter for GNNs.}
	\vspace{-10pt}
	\label{fig:eval:gcn-pyg}
\end{figure}

\textbf{Comparison with PyG.} PyG~\cite{pyg} provides an example code for GCN. We repeat running GCN on different graphs with different model settings on PyG 
and report the total reduction of CUDA time in Fig.~\ref{fig:eval:gcn-pyg}. Improvements on PyG is more obvious than on DGL. This is because the general MessagePassing API in PyG explicitly generates `message' on all edges and then performs reduction, while DGL fuses the two phases into one SpMM kernel. To summarize, integrating GE-SpMM brings up to 3.67$\times$ and 2.10$\times$ CUDA time reduction on two GPUs. 

\subsubsection{GNNs based on SpMM-like Operators} 

GE-SpMM is intended to accelerate a class of SpMM-like operators that is not yet supported by cuSPARSE. One good example is the pooling operation in GraphSAGE-pool~\cite{sage}, where each vertex aggregates the features of its neighbors by taking maximum. cuSPARSE does not provide this operation. Future GNN models may also use customized reduction functions for pooling, and relying on cuSPARSE is not flexible to these user-defined operations. It is not hard to generalize GE-SpMM to support SpMM-like operations, since these operations follow similar memory access patterns and the code is almost the same. We implement a GE-SpMM kernel for the aggregation step in GraphSAGE-pool. Our kernel brings up to 1.14$\times$ acceleration on the CUDA time of GraphSAGE-pool training on Pubmed graph. Note that for SpMM-like operation only, GE-SpMM can achieve 2.39$\times$ to 6.15$\times$ speedup over DGL's SpMM-like kernel. This shows the value of this work in supporting flexible, user-defined SpMM-like operation in new GNN models, which are not covered by the vendor library, and achieve even better performance. 

\begin{table}[h]
    \centering
    \caption{Summary of GraphSAGE-pool CUDA time reduction on DGL.}
    \begin{tabular}{c|c|c|c|c}
    
        \hline
         & \multicolumn{2}{c}{GTX 1080Ti} & \multicolumn{
         2}{ |c}{RTX 2080}\\
         \hline
         (\#layer, \#feature) & SpMM-like & Total & SpMM-like & total\\
         \hline
         (1,16) & 2.89 & 1.10 & 3.24 & 1.11 \\
         (1,64) & 3.92 & 1.14 & 3.44 & 1.11 \\
         (1,256) & 4.04 & 1.14 & 3.46 & 1.09 \\
         (2,16) & 2.39 & 1.11 & 3.03 & 1.09 \\
         (2,64) & 3.09 & 1.09 & 3.37 & 1.11 \\
         (2,256) & 6.15 & 1.12 & 3.51 & 1.09 \\
         \hline
    \end{tabular}
    \label{tab:my_label}
\end{table}

\section{Conclusion} \label{sec:conc}
In this paper, we propose an efficient CSR-based SpMM design, GE-SpMM, for Graph Neural Network applications on GPUs. GE-SpMM considers requirements by GNN applications, including no preprocessing and SpMM-like operation requirements. GE-SpMM introduces two techniques: Coalesced Row Caching and Coarse-grained Warp Merging, to improve the efficiency of global data access, leading to 1.25$\times$ and 1.51$\times$ speedup, respectively. By adopting these optimizations, GE-SpMM achieves up to 1.41$\times$ speedup over Nvidia cuSPARSE~\cite{cusparse} and up to 1.81$\times$ over GraphBLAST~\cite{design}. We also embed GE-SpMM in GNN frameworks (e.g., DGL, PyG) and achieve significant CUDA time reduction.

\bibliography{references.bib}
\bibliographystyle{IEEEtran}

\end{document}